\documentclass{epl}

\title{Mode Coupling relaxation scenario in a confined glass \\ former}
\shorttitle{Mode Coupling relaxation scenario in a confined ...}
\author{P. Gallo \and R. Pellarin \and M. Rovere}
\shortauthor{P. Gallo et al.}
\institute{Dipartimento di Fisica ``E. Amaldi'', 
Universit{\`a} ``Roma Tre'', \\ 
and Istituto Nazionale per la Fisica della Materia,  Unit{\`a} di Ricerca
Roma Tre, \\ Via della Vasca Navale 84, I-00146 Roma, Italy
}

\begin{document}

\maketitle

\begin{abstract}
Molecular dynamics simulations of a  Lennard-Jones binary mixture
confined in a disordered array of soft spheres are presented.
The single particle dynamical behavior of the glass former is
examined upon supercooling. Predictions of
mode coupling theory are satisfied by the confined liquid. 
Estimates of the crossover 
temperature are obtained by power law fit
to the diffusion coefficients and relaxation times of the late $\alpha$ region.
The $b$ exponent of the von Schweidler law is also evaluated.
Similarly to the bulk, different values of the exponent $\gamma$
are extracted from the power law fit to 
the diffusion coefficients and relaxation
times. 
\end{abstract}


The glass transition scenario in confined liquids is a field
of rapidly growing interest because of its connection with 
relevant problems in technology and biology~\cite{grenoble}. 
One of the momentous questions in this field is whether 
and to what extent theories successfully used 
for investigating the glass transition in the bulk
retain their validity upon confinement.
Experiments  
\cite{Daoukaki,Mckenna,kremer1,kremer2,melni,enza,zanotti1,zanotti2} 
show a rather diversified phenomenology.  Although
in most cases the process of vitrification
of a glass former upon supercooling is not suppressed by confinement,
substantial modifications with respect to the bulk 
behavior are almost always observed

In the theoretical framework of the glass transition 
of bulk liquids a relevant role is played 
by Mode Coupling Theory (MCT)~\cite{goetze}. This theory is able to predict 
the approach of the supercooled liquid~\cite{pablo}
that is undergoing a calorimetric glass transition to a temperature
$T_C$ that signals the ideal crossover from a regime where the breaking and
the formation of cages is the mechanism that ensures the system to be 
ergotic to a regime where cages are frozen and only hopping processes
restore ergodicity. 

The approach to the glass transition of binary mixtures in the bulk phase
has been the subject of numerous studies, see for example
\cite{bernu2,goran,fujiara}. 
In particular a numerical test of the MCT predictions 
has been successfully carried out for a Lennard Jones binary
mixture (LJBM)~\cite{kob}.

Real porous solids are often disordered, they are made of
an interconnected network of voids of various size and
shapes. In order to investigate the influence of a disordered microstructure
on the glass transition scenario we assume for the solid the same model
used in the past both in computer simulation and in
statistical mechanical theories to study the modifications
of the phase diagram in confined fluids: the solid is treated as
a collection of particles forming a rigid 
disordered structure~\cite{rosinberg,monson,page}. 
In this way the fluid is embedded in an off-lattice matrix 
of rigid ``obstacles''. This model is in particular representative
of amorphous materials like silica xerogels.

In this letter we demonstrate that the main predictions of MCT
are satisfied for a glass former liquid
confined in this disordered medium. 

We introduced into a rigid disordered matrix of 16 soft spheres  
the LJBM of 800 A and 200 B particles
as defined in ref.~\cite{kob}. The use of a binary mixture
is strictly necessary in the bulk to avoid cristallization
\cite{bernu1}. Confinment 
is expected to inhibit cristallization also in a one component fluid
instead. We did however choose to use the LJBM of  ref.~\cite{kob}
for a comparison with the bulk. The parameters of the LJBM are 
$\epsilon_{AA}=1$, $\sigma_{AA}=1$, $\epsilon_{BB}=0.5$,
$\sigma_{BB}=0.88$, $\epsilon_{AB}=1.5$, and $\sigma_{AB}=0.8$.
In the following energy will be given in units of
$\epsilon_{AA}$, temperature in units of 
$\epsilon_{AA}/K_B$, length in units $\sigma_{AA}$ and time in units of
$(m\sigma_{AA}^2/(48\epsilon_{AA}))^{1/2}$.
The Lennard-Jones particles interact 
with the soft spheres with a repulsive potential:
\begin{equation}
V(r)= 4 \epsilon\left({\sigma_{S}\over r}\right)^{12}
\end{equation}
with
$\epsilon_{SA}=0.32$, $\sigma_{SA}=3$, 
$\epsilon_{SB}=0.22$, $\sigma_{SB}=2.94$.
Parameters have been chosen in order to obtain
a strongly confined system where most of the LJ particles
are interfacial.  
The potential is truncated at $r^c_{ij}=2.5 \sigma_{ij}$.
We conducted the simulations in the microcanonical ensemble
in a cubic box of length $12.6$  with periodic boundary conditions.
The equations of motion were solved by the velocity Verlet algorithm.

The system was equilibrated at different reduced temperatures 
via a velocity rescaling procedure starting from $T=5.0$. 
It was then progressively cooled and equilibrated at the following 
temperatures $T=4.0$, $3.0$, $2.0$, $1.0$. Below  $T=1.0$,
after equilibration, production runs were done for the following
temperatures: $T=0.80$, $0.58$, $0.48$, $0.43$, $0.41$,
$0.39$ and $0.37$. 
The timestep used for $T\ge 1$ was $0.01$ and  for  $T<1$ was $0.02$. 
For the lowest temperature investigated a production run of ten million
timesteps was performed. 
We verified that the results presented in this letter
do not depend on the specific choice of the disordered matrix
by running MD simulations for other two different configurations
of the disordered matrix of soft spheres.
Both thermodynamics and dynamics appeared the same for the three systems, 
in particular deviations among the relaxation times 
extracted at a given temperature from the density correlators
of different configurations are within $2$\%.

A snapshot of the system under investigation is 
shown for a low temperature in Fig.~1. 
From the picture it is evident that we are dealing with a strongly
confined system. 

No phase separation of A and B particles 
is detected during the cooling process through the calculation of the
pair correlation functions (not shown)~\cite{noi}.

In the following we concentrate on the single particle dynamics
of A and B particles.
In Fig.~2 the Mean Square Displacement
(MSD) for the A particles is shown for all 
the temperatures investigated. It displays the typical
behavior of a system approaching the crossover temperature of MCT.
Upon lowering the temperature it appears an intermediate time region
where the MSD flattens due to the well known cage effect.
Similar behavior is found for the B particles (not shown).
From the slope of the MSD, once the particle enters the Brownian 
diffusive regime,
it is possible to extract the diffusion coefficient which is predicted by MCT
to have a power law behavior:
\begin{equation}
D\sim(T-T_C)^\gamma
\label{pl1}
\end{equation}
In the inset of Fig.~2 we show the diffusion coefficients extracted 
at the different temperatures from the MSD both for A and B particles 
together with the fit to Eq.~\ref{pl1}.
For A particles $T_C=0.343$ and $\gamma=1.60$ and for 
B particles  $T_C=0.343$ and $\gamma=1.69$ are found. 
We obtain the same
value of $T_C$ for both species, as predicted by MCT, but 
slightly different values of 
$\gamma$. This discrepancy was also found in the bulk~\cite{kob}.

In Fig.~3 we show the density-density self correlation function
$F_S(Q,t)$ at the peak of the AA structure factor, which for
this system is $Q\sigma_{AA}\simeq 7.07$ as shown in the inset of 
Fig.~4. 
We calculated $F_S(Q,t)$ also for B particles at the peak 
of the BB structure factor $Q\sigma_{BB}=5.90$ (not shown).
As already displayed by the MSD for higher temperatures 
the function does show a single relaxation timescale  while 
upon cooling we observe the shouldering of the relaxation law 
that corresponds to the onset of the slow $\alpha$-relaxation regime in the
system.
The late part of the $\alpha$ relaxation region is found 
in most glass formers to have
the analytical shape of the Kohlrausch-William-Watts, KWW, function:
\begin{equation}
\phi_Q (t)=f_Qe^{-(t/\tau)^\beta} 
\label{strexp}
\end{equation}
where $f_Q$ is the height of the plateau and $\tau$ is the
late $\alpha$ relaxation time.
The fit to this formula are also shown in Fig.~3. The agreement is
very satisfactory. In the lower inset we report the $\beta$ and $f_Q$ 
values extracted
for the A and B particles from the fit. 
In the upper inset the $\tau$ values are shown together
with the fit to the MCT power law:
\begin{equation}
\tau\sim(T-T_c)^{-\gamma}
\label{taugamma}
\end{equation}
For A particles $T_C=0.343$ and $\gamma=2.90$ and for 
B particles  $T_C=0.343$ and $\gamma=2.89$ are extracted.
As found in the bulk the values of $T_C$ predicted from
$D$ and $\tau$ are similar while the $\gamma$ are different.

We now move to the so called von Schweidler test. 
MCT predicts that the region of departure from the plateau
of the density correlator in the $Q,t$ space,
the late $\beta$ relaxation region, behaves according to the following 
power law:
\begin{equation}
\phi_Q (t)=f^c_Q-h_Q(t/\tau)^b 
\label{VS}
\end{equation}
where $\tau$ is a characteristic time of the system,
$b$ a system dependent exponent, $f^c_Q$ is the height of
the plateau at the temperature $T_C$ and $h_Q$ an amplitude that
does not depend on time.
In Fig.~4 we show the $F_S(Q,t)$ of Fig.~3 now rescaled 
according to the $\tau$ values extracted from the fit to Eq.~\ref{strexp}.
We note that all the curves collapse into a single master curve
as predicted by MCT. The deviations
from the master curve appear to be mostly due to overshots clearly visible 
in our correlators
and also present in the bulk that tend to mask the MCT predicted
behavior. This kind 
of oscillations appear to be more marked upon confinement.
In the same figure the fit of the master curve to Eq.~\ref{VS}
is also reported. We obtain $b=0.355$ and $f^c=0.72$ for A particles, 
$b=0.35$ and $f^c = 0.785$ for B particles.

MCT also predicts that the region of approach to the plateau
is given by another power law characterized by  
an exponent $a$, also called the critical exponent. $a$ cannot
be directly extracted by our data due to the oscillations that
start appearing in the correlator upon cooling, as also
evident in Fig.~3. 
 
With the values of $b$ and the $\gamma$ extracted by Eq.~\ref{taugamma}
and by means of the MCT relationship:
\begin{equation}
\gamma ={1\over{2a}}+{1\over{2b}}
\end{equation}
we obtain $a=0.335$ for the A particles and 
$a=0.342$ for the B particles.
The values of $\gamma$, $a$ and $b$ are
well in the range predicted by the theory. 
The exponents $a$ and $b$ both allow an evaluation of the parameter
$\lambda$ of MCT, also known as the exponent parameter, $1/2 < \lambda< 1$,
by means of the MCT formula:
\begin{equation} 
\lambda= \frac{\left[ \Gamma  \left( 1-a \right) \right]^2} {\Gamma 
\left( 1-2a \right)}=
\frac{\left[ \Gamma  \left( 1+b \right) \right]^2}{\Gamma \left( 1+2b \right)}
\end{equation}
where $\Gamma (x)$ is the $\Gamma$ function.

The system dependent parameter $\lambda$ governs the behaviour of 
$f_Q$ close to $f^c_Q$ below $T_C$. 
We have a difference of $22$\%
for the values of $\lambda$ extracted from $b$ and $a$ for both A and B
species which is larger with respect to the value $5$\% found 
for the bulk~\cite{kob}. 
In our case we observe that, due to the inevitable oscillations
of the correlators, the fit of the von Schweidler law that allow the
determination of the exponent $b$ is affected by a larger uncertainty
with respect to the bulk case.
Therefore the best estimate we can obtain for the parameter $\lambda$ 
of this system is: $\lambda=0.8\pm0.1$ where we are attributing to
$\lambda$ the maximum error coming form the semidispersion of the 
values obtained.
We note that for a more precise determination of the exponent $b$
also the term next to the leading order has to be taken into account
in eq.\ref{VS} for the fit and at the same time several $Q$ values should be 
tested simultaneously \cite{fs}. This analysis is in progress and will
be reported in a subsequent paper \cite{noi}.

In conclusion we performed a MD simulation on a LJBM
confined in a disordered matrix of soft spheres.
Our results provide insight into the microscopic behavior
of this confined fluid when cooled. Since no cristalline equilibrium 
solid can be produced in the confined matrix
we are not able to individuate the freezing temperature
below which the liquid can be considered supercooled as it is done in 
the bulk \cite{fujiara}.
None-the-less most of liquids start displaying the shouldering of the
relaxation laws below the freezing temperature, we can therefore 
reasonably consider the fluid supercooled for temperatures below $T=0.5$.  
We have shown that MCT is able to rationalize the dynamical behavior
of the system in spite of the fact that this theory is formulated
for bulk liquids.
In particular both the diffusion coefficients extracted from the slope of
the MSD and the relaxation times extracted from the fit to the KWW
function to the late part of the $\alpha$ relaxation region of the 
intermediate scattering function allow to estimate the crossover
temperature $T_C=0.343$ of the confined LJBM.
The rescaled intermediate scattering functions define a master curve
that can be fitted to the von Schweidler law from which the $b$
exponent is extracted. 
From the values of $a$ and $b$ the exponent parameter $\lambda$ 
is also evaluated.

An MCT test with MD on confined liquids has been so far carried out 
to the best of our knowledge 
only for water confined in a silica nanopore \cite{jcp-noi}. 
In that case due to the different kind of confining medium and
to the different liquid the MCT behaviour could be extracted only with
a layer analysis of the density correlators.
For the system presented here this is not necessary.
Besides the obvious differences in the
confining geometry and liquid another 
important difference between the two systems is that the LJBM is
in a situation of strong confinement. In fact given the paramenter
of the soft spheres only roughly 1/3 of the total volume is accessible 
for the mixture and only few layers of LJBM can reside among 
the spheres as it is also evident from Fig.~1. In the case of water 
many layers can reside in a nanopore.

Although our system
is not directly comparable to any experimental system,
simple fluids adsorbed in silica xerogels are likely to display similar 
behavior~\cite{rosinberg,monson,page}. Experimental MCT tests
on such systems would therefore be very valuable to assess the validity
of our findings. 

\acknowledgments
We acknowledge the financial support of the
G Section of the INFM. We all thank G. Pastore and
M.R. thanks R. Evans for interesting discussions.

\newpage
\begin{figure}
\onefigure[scale=0.75]{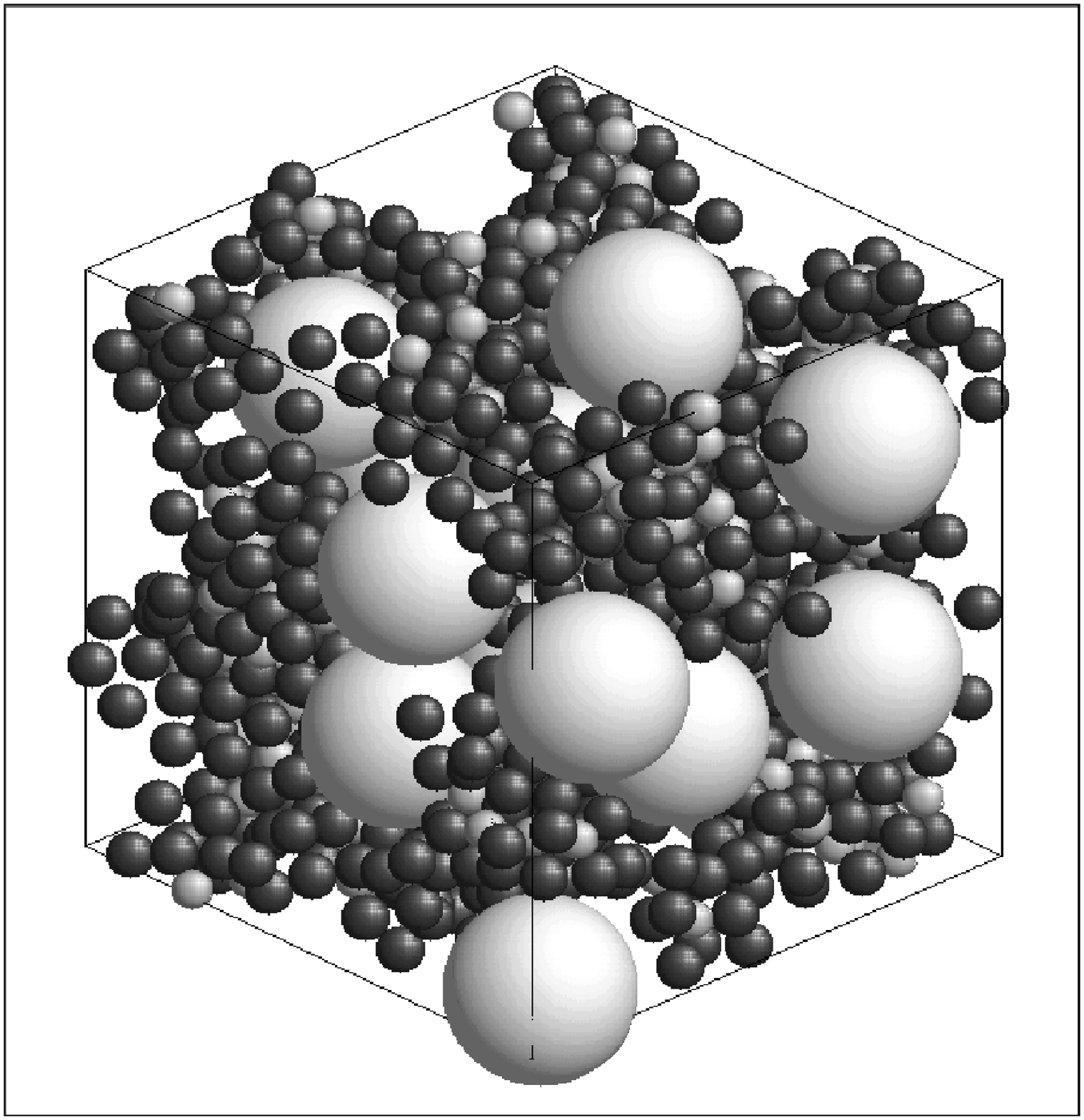}
\caption{Snapshot of the system simulated at $T=0.4$. 
Larger spheres represent the confining system, 
dark-gray particles the A type, light-gray the B type particles of the 
LJBM.}
\label{fig:1}
\end{figure} 

\begin{figure}
\onefigure{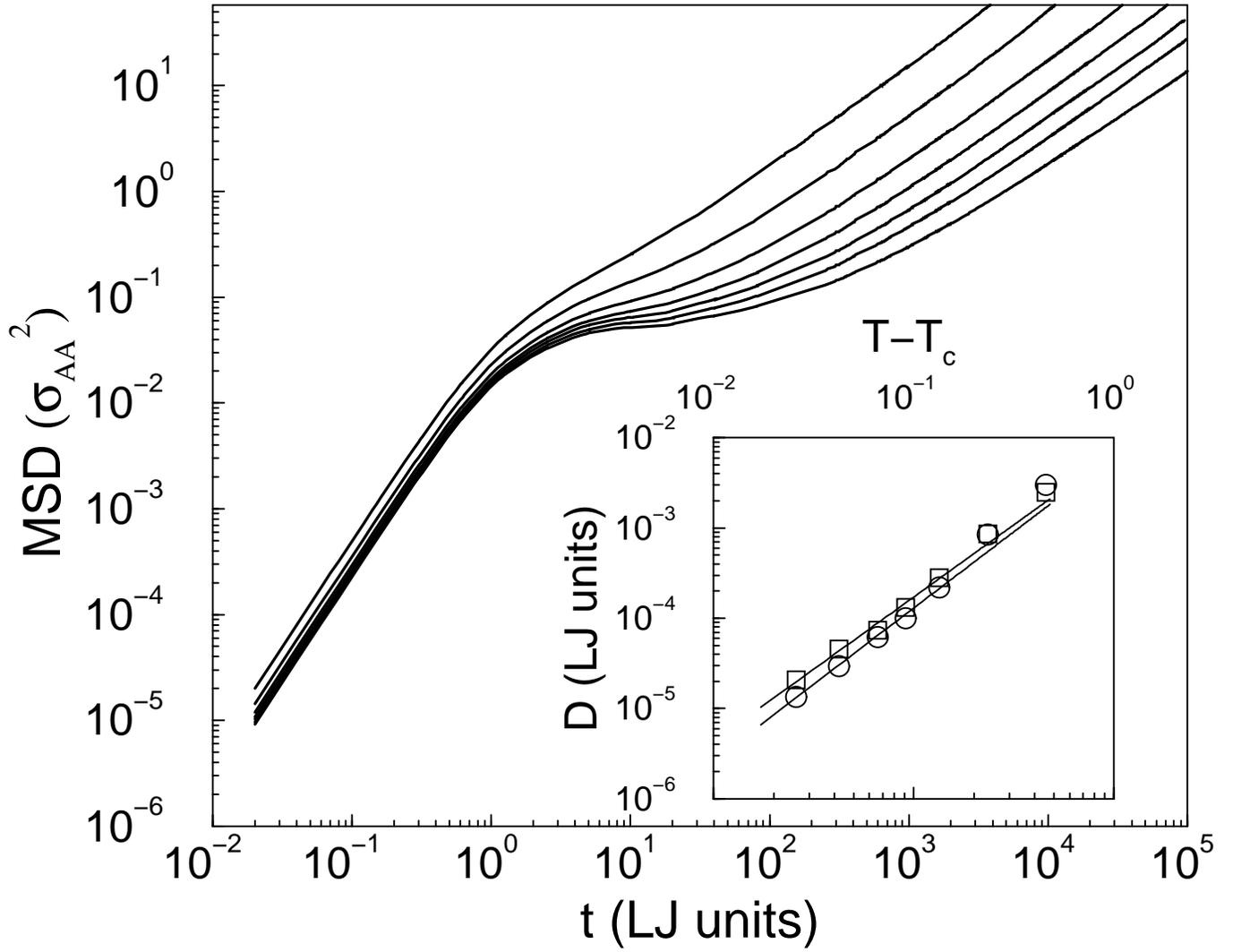}
\caption{Mean square displacement of A particles for 
$T=0.80$, $0.58$, $0.48$, $0.43$, $0.41$, $0.39$, $0.37$.
Curves on the top correspond to higher temperatures. 
In the inset we show the diffusion coefficient D as a function 
of temperature for A (squares)
and B (circles) particles together with the fit (continuous lines) to the 
power law of Eq.~\protect\ref{pl1}.
For A particles $T_C=0.343$ and $\gamma=1.60$, for 
B particles  $T_C=0.343$ and $\gamma=1.69$.}
\label{fig:2}
\end{figure} 

\begin{figure}
\onefigure{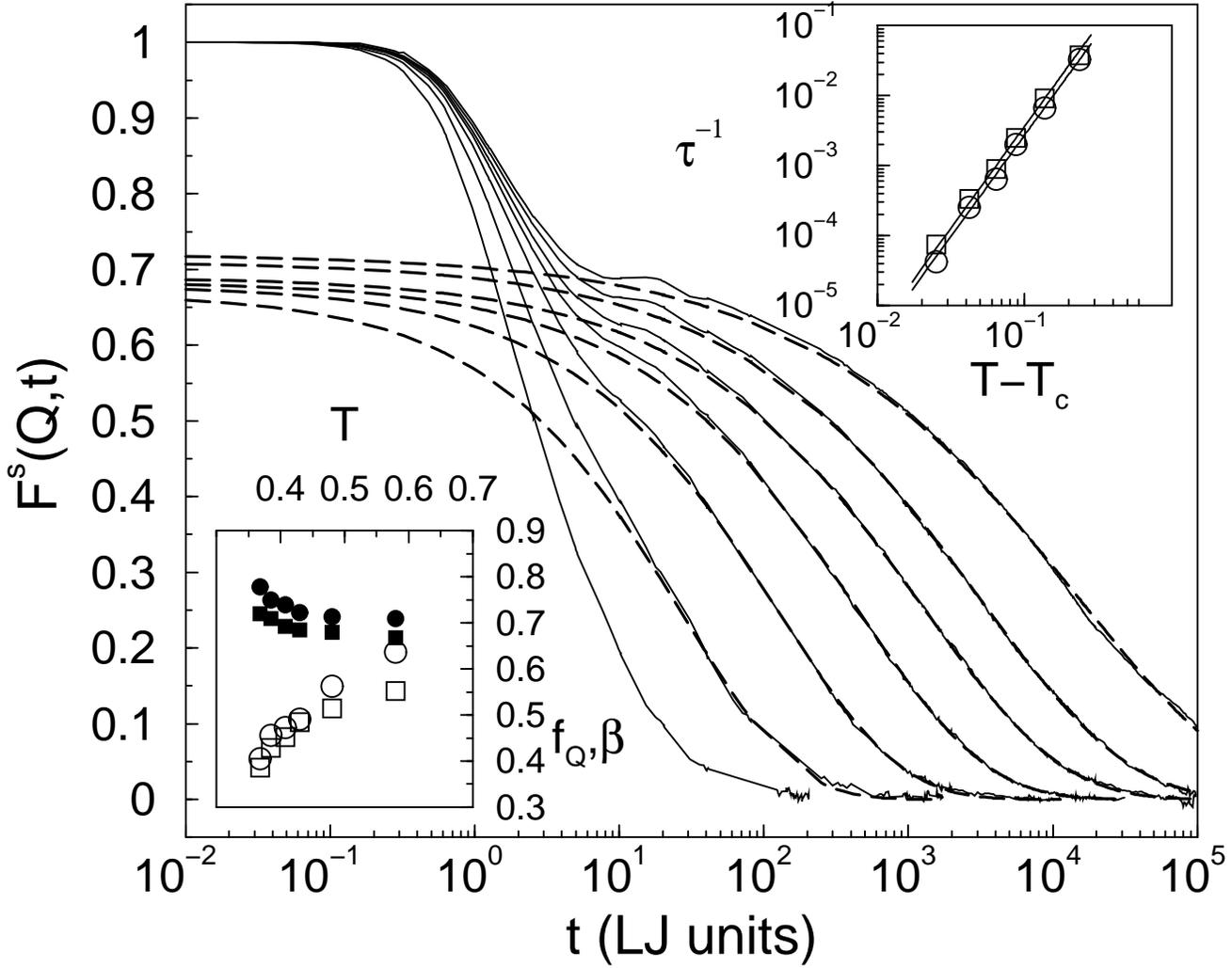}
\caption{Self part of the intermediate scattering function at
the peak of the structure factor $Q_{MAX}=7.07 \sigma_{AA}^{-1}$ for 
A particles (continuous line) for the same temperatures as in 
Fig.~{\ref{fig:2}}.  
Curves on the top correspond to lower temperatures.
The dashed curves are the KWW fits. In the lower inset we show  
the non-ergodicity parameter $f_Q$ (filled symbols) and 
the stretching exponent $\beta$ (empty symbols)
for A (squares) and B (circles) particles,
as extracted from the fits to the KWW. 
In the upper inset the power law fit (continuous lines)
for the relaxation time is shown
for A (squares) and B (circles) particles. For A particles 
$T_C=0.343$ and $\gamma=2.90$, for 
B particles  $T_C=0.343$ and $\gamma=2.89$.}
\label{fig:3}
\end{figure} 

\begin{figure}
\onefigure{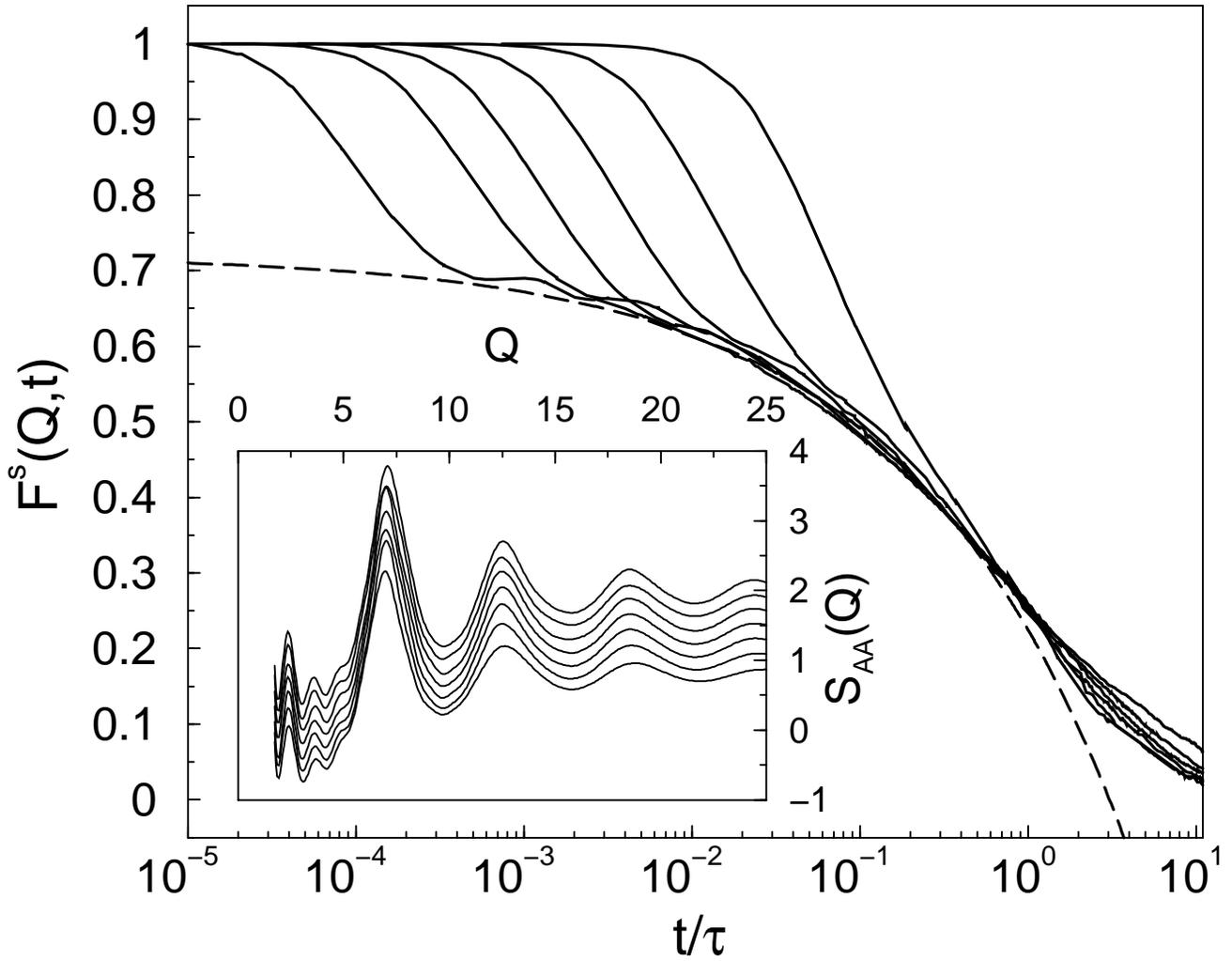}
\caption{In the main frame continuous lines are the 
time-rescaled intermediate scattering functions of Fig.~\ref{fig:3}.
The dashed line is the fit to the master curve according to the
von Schweidler law.  We obtain $b=0.355$ and $f^c_Q=0.72$ for A particles, 
$b=0.35$ and $f^c_Q = 0.785$ for B particles (not shown).
In the inset the static structure factor of A particles
is shown for all the temperatures investigated. For the sake of
clarity lower temperatures have been upward shifted. 
The first peak is located at $Q_{MAX}=7.07 \sigma_{AA}^{-1}$. }
\label{fig:4}
\end{figure} 

\end{document}